\documentclass[bib]{ba}
\usepackage[none]{hyphenat}
\usepackage{natbib}
\usepackage{graphicx}
\usepackage{color}

\usepackage{bayes}
\usepackage{float}
\usepackage{caption}
\usepackage{placeins}
\usepackage{enumitem}
\usepackage{bm, amsmath, amssymb}
\usepackage[titletoc]{appendix}
\usepackage{verbatim} 

\newtheorem{theorem}{Theorem}[section]

\newtheorem{proposition}[theorem]{Proposition}

  \newcommand{\qed}{\nobreak \ifvmode \relax \else
      \ifdim\lastskip<1.5em \hskip-\lastskip
      \hskip1.5em plus0em minus0.5em \fi \nobreak
      \vrule height0.75em width0.5em depth0.25em\fi}
\newcommand{\bYi}{{\bm Y}_{\bm{t}_i}}
\newcommand{\bZi}{{\bm Z}_{\bm{t}_i}}
\newcommand{\bZis}{\bm{Z}_{{\bf t}_i^{\ast}}}
\newcommand{\bti}{\bm{t}_i}

\begin{document}

{\sloppy

\inserttype[ba0001]{article}
\renewcommand{\thefootnote}{\fnsymbol{footnote}}
\author{Jingjing Yang, Hongxiao Zhu,  Dennis D.~Cox }{

 \fnms{Jingjing}
 \snm{Yang}
 \footnotemark[3]\ead{yjingj@umich.edu},

  \fnms{Hongxiao}
  \snm{Zhu}
  \footnotemark[2]\ead{hongxiao@vt.edu},
  
    \fnms{Taeryon}
  \snm{Choi}
  \footnotemark[4]\ead{trchoi@korea.ac.kr},
  
  \fnms{Dennis D.}
  \snm{Cox}
  \footnotemark[1]\ead{dcox@rice.edu}

}

\title[]{Smoothing and mean-covariance estimation of functional data with a Bayesian hierarchical model
}

\maketitle
\footnotetext[3]{
 Department of Biostatistics, University of Michigan, Ann Arbor, MI 48109, USA, yjingj@umich.edu
}
\footnotetext[1]{
 Department of Statistics, Rice University, Houston, TX 77005, USA, dcox@rice.edu
}
\footnotetext[2]{
 Department of Statistics, Virginia Tech, Blacksburg, VA 24061, USA, hongxiao@vt.edu
}

\footnotetext[4]{
Department of Statistics, Korea University, Seoul 136-701, Republic of Korea, trchoi@korea.ac.kr
}

\renewcommand{\thefootnote}{\arabic{footnote}}

\begin{abstract}
Functional data, with basic observational units being functions (e.g., curves, surfaces) varying over a continuum, are frequently encountered in various applications. While many statistical tools have been developed for functional data analysis, the issue of smoothing all functional observations simultaneously is less studied. Existing methods often focus on smoothing each individual function separately,  at the risk of removing important systematic patterns common across functions. We propose a nonparametric Bayesian approach to smooth all functional observations simultaneously and nonparametrically. In the proposed approach, we assume that the functional observations are independent Gaussian processes subject to a common level of measurement errors, enabling the borrowing of strength across all observations. 
Unlike most Gaussian process regression models that rely on pre-specified structures for the covariance kernel, we adopt a hierarchical framework by assuming a Gaussian process prior for the mean function and an Inverse-Wishart process prior for the covariance  function. These prior assumptions induce an automatic mean-covariance estimation in the posterior inference in addition to the simultaneous smoothing of all observations. Such a hierarchical framework is flexible enough to incorporate functional data with different characteristics, including data measured on either common or uncommon grids, and data with either stationary or nonstationary covariance structures. Simulations and real data analysis demonstrate that, in comparison with alternative methods, the proposed Bayesian approach achieves better smoothing accuracy and comparable mean-covariance estimation results. Furthermore, it can successfully retain the systematic patterns in the functional observations that are usually neglected by the existing functional data analyses based on individual-curve smoothing.


\end{abstract}

\section{Introduction}
As more digital data are being collected in modern experiments, considerable efforts have been made to process and analyze {\it functional data} --- data that are realizations of random functions varying over a continuum such as a region of time or a range of wavelengths \citep{ramsay2002applied}. Since \citet{ramsay1991some} first coined the term ``functional data analysis'' (FDA), numerous papers have been published on the FDA theory, methods and applications, making it one of the most active research areas in statistics. One salient feature of functional data is that, although the underlying functions are often continuous and smooth, data can only be collected discretely, which often produces measurement errors. 
It is possible to treat functional data as fully observed when the measurements are dense and the noise level is low \citep{hall2001, cardot2003spline, Zhu200901, Zhu2010}. However, in most circumstances, it is necessary to convert the discrete measurements to functions that can be evaluated on any subset of the domain, referred to as smoothing. Smoothing reduces the effect of measurement errors, and better smoothing leads to less biased results in further analyses (see, e.g., \citet{hitchcock2006improved}).


Existing approaches for smoothing functional data can be divided into three overlapping categories (see, e.g., \cite{ramsay2006functional}):  (1) methods based on representations with basis functions such as B-splines, Fourier series, and wavelets; (2) nonparametric smoothing with kernels; and (3) roughness-penalty-based methods such as smoothing spline. 
Although these existing smoothing methods are computationally tractable and generally available in standard statistical softwares, they mainly focus on smoothing each individual curve separately rather than smoothing all curves at the same time, thus failing to borrow strength across all observations.  Such a failure may cause the following problems: (1) The order of smoothness of each curve may vary across all functional observations even if they are assumed to be independent and identically distributed (i.i.d.). This problem is severe when functional data are sparse and measured on uncommon grids. (2) The systematic patterns shared by all observations could be blurred or wiped out. This problem becomes more severe when the signal-to-noise-ratio (SNR) is low.

In these regards, we propose a model-based approach for smoothing all functional observations simultaneously and nonparametrically. In particular, we assume that all functional observations are
i.i.d.~Gaussian processes (GPs) subject to a common level of 
additive measurement errors, which facilitates the borrowing of strength across all data
and the simultaneous smoothing based on the hierarchical GP structures. 
The GP prior is a nonparametric prior widely used in Bayesian nonparametric regression, spatial modeling, and machine learning (see, e.g., \cite{Rasmussen2006}, \cite{banerjee2008gaussian}, \cite{Banerjee2012}, and \cite{banerjee2014hierarchical}). Recently, GP prior has also been increasingly  employed in Bayesian FDA. For instance, GP has been used as the prior for functional batch effects \citep{kaufman2010bayesian} and as the base measure for a Dirichlet process prior \citep{nguyen2014bayesian} in functional ANOVA. GP regression has also been used in FDA (see, e.g., \cite{shi2011}, \cite{Shietal12}, and \cite{WangShi14}),  providing a nonparametric linkage between a functional predictor and a functional response.
Despite the wide applicability of GP,
most existing approaches assume pre-specified parametric structures for the covariance kernel such as squared exponential or Mat{\'e}rn, and the parameters in the kernel function often need to be pre-determined, which causes a lack of flexibility and introduces potential biases.

We contrast our proposed approach with existing GP methods for exploiting a Bayesian hierarchical model with a GP prior for the mean curve and an Inverse-Wishart process (IWP) prior for the covariance surface. By assuming the IWP prior, our approach avoids possible biases caused by the mis-specification of the covariance kernel as commonly used in the existing methods. Furthermore, our method has the flexibility of allowing the scale parameter of IWP to 
adjust a smooth covariance surface to be either stationary or nonstationary, which provides more flexible posterior covariance estimation as well as more accurate smoothing of functional observations

In addition, through borrowing strength across all curves, our approach facilitates the smoothing of sparse functional data measured on uncommon grids. An individual curve contains much less information about the functional observation 
when it is observed at very few grid points. Smoothing individual curves one-at-a-time is extremely difficult in this case. With the assumption that the functional observations share the common  mean and covariance, our Bayesian approach can simultaneously smooth all curves on a pooled, dense grid, by treating functional values at unobserved grid points as parameters (or missing values) and updating them during a Markov Chain Monte Carlo (MCMC) procedure.

Finally, our {proposed Bayesian inferential procedure also produces} mean and covariance estimation as by-products. Functional mean-covariance estimation has already been extensively investigated in the literature. Existing approaches include the sample estimation \citep{ramsay2006functional}, the roughness-penalty-based method \citep[page 22]{RiceSilverman1991},  and the Principal Analysis by Conditional Expectation (PACE) method with local polynomial smoothing \citep{Yao2005}. Regarding the mean-covariance estimation, our proposed method can be considered as a Bayesian counterpart of PACE. PACE aims to estimate eigenvalues and eigenfunctions of the covariance surface with irregularly spaced longitudinal data \citep{Yao2005,yao2005functional,muller2005,Leng2006}. It also produces mean-covariance estimates as by-products. The main idea of PACE is to first formulate a ``raw" covariance using pooled sparse longitudinal measurements and then apply a two-dimensional local polynomial smoother to estimate the covariance. Smoothing based on the pooled raw data has the effect of borrowing strength from all data; performing local polynomial smoothing has a similar effect as performing Bayesian smoothing with a GP prior. Therefore, the mean-covariance estimates produced by our approach should be comparable with the ones produced by PACE. In comparison with PACE,  however, our Bayesian approach offers the benefit of full posterior inference for all latent processes and parameters, such as credible intervals and simultaneous credible bands for the mean curve and the covariance surface. These uncertainties are hard to quantify in a frequentist approach such as PACE. Furthermore, our method avoids the selection of optimal bandwidths for local polynomial smoothing as required in PACE, which can be sensitive to  different selection methods.

Accordingly, the main contributions of our proposed approach can be summarized as follows:
(1) It facilitates simultaneous smoothing of all functional observations.  (2) It preserves systematic patterns that are common across all curves. 
(3) It avoids the potential bias caused by the mis-specification of the covariance function.  
(4) It is flexible enough to accommodate functional data with common or uncommon grids, dense or sparse measurements, as well as stationarity or nonstationarity. (5) It results in mean-covariance estimates as by-products, with full Bayesian characterization of the uncertainties.


To demonstrate the performance of our proposed method, we considered simulation studies with synthetic data 
as well as two case studies with real spectroscopy and metabolic data. The results are compared with alternative methods based on single-curve smoothing (e.g., smoothing spline and kernel smoothing) as well as PACE. In simulation studies with noisy functional data based on either stationary or nonstationary covariances, we find that our proposed approach recovers the true signals more accurately than those single-curve smoothing methods. In real case studies, we find that our method retains systematic patterns that were wiped out by the single-curve smoothing methods. Regarding the mean-covariance estimation, our proposed approach is comparable with the PACE method as expected.

The remainder of this article is organized as follows: In Section \ref{Bayes:model}, we outline the details of the proposed Bayesian hierarchical model as well as the heuristic empirical Bayes approach used to determine hyperparameters.  In Section \ref{Bayes:mcmc}, we describe the MCMC algorithm for posterior inference.  Results of simulations and real case studies are presented in Sections \ref{Bayes:sim} and \ref{Bayes:real}, respectively. Discussion is provided in Section \ref{Bayes:disc}.

\section{Unified Bayesian hierarchical model}
\label{Bayes:model}

\subsection{Model description}
\label{Bayes:assump}

Suppose that the functional data contain $n$ independent trajectories,
denoted by $\{Y_i(\cdot);\, i = 1, 2, \cdots, n\}$, and
the $i^{th}$ trajectory has $p_i$ measurements
on  grid $\bm{t}_i = \{t_{i1}, \cdots, t_{ip_i}\}$.
Further assume that the $i^{th}$ trajectory $Y_i(\cdot)$ depends on an underlying
GP $Z_i(t)$, $t\in \mathcal{T}$, through the following model:
\begin{equation}
  Y_i(t_{ij}) = Z_i(t_{ij}) + \epsilon_{ij},
  \quad t_{ij} \in \mathcal{T},
  \quad i = 1, \cdots, n, \quad j = 1, \cdots, p_i,
  \label{bayes:eq2.1}
\end{equation}
where $\{Z_i(\cdot)\}$ are i.i.d. GPs with mean function $\mu(\cdot)$ and covariance kernel $\Sigma(\cdot, \cdot)$, denoted by $Z_i \sim GP(\mu, \Sigma)$. The covariance kernel satisfies $\Sigma(s, t) = E[(Z_i(s) - \mu(s))(Z_i(t) - \mu(t))]$, $\forall s, t \in \mathcal{T}$.
The error terms $\{\epsilon_{ij}\}$ are assumed to be i.i.d.~random normal variables, i.e., $\epsilon_{ij}\sim N(0, \sigma^2_{\epsilon})$, and independent of $\{Z_i(\cdot)\}$. We assume the following priors
for the model parameters $\sigma_{\epsilon}^2$, $\mu(\cdot)$, and $\Sigma(\cdot, \cdot)$:
\begin{eqnarray}
\sigma_{\epsilon}^2 \sim \mbox{Inverse-Gamma}(a_{\epsilon}, b_{\epsilon}), \quad (\mu\mid \Sigma) \sim GP\left(\mu_0, \frac{1}{c}\Sigma\right), \quad \Sigma \sim \mbox{IWP}(\delta, \Psi),
\label{eq:prior}
\end{eqnarray}
where $\mbox{Inverse-Gamma}(a_{\epsilon}, b_{\epsilon})$ denotes an Inverse Gamma distribution with shape parameter $a_{\epsilon}$ and scale parameter $b_{\epsilon}$. The scaling parameter $c$ in the GP prior satisfies $c>0$. Here $\mbox{IWP}(\delta, \Psi)$ denotes the IWP with shape parameter $\delta$ 
 and scaling parameter $\Psi$. An IWP is defined such that, on any finite grid
$\bm{t} = \{t_1, t_2, \cdots, t_p\}$ with $p$ points, the projection
$\Sigma(\bm{t}, \bm{t})$ is Inverse-Wishart distributed, i.e., $\Sigma(\bm{t}, \bm{t}) \sim \mbox{IW}(\delta, \Psi(\bm{t}, \bm{t}))$. Here the
 Inverse-Wishart distribution follows the parametrization of  \citet{dawid1981some}. In particular,
a symmetric and positive definite matrix $\Sigma(\bm{t}, \bm{t})$ is said to be
$\mbox{IW}(\delta, \Psi(\bm{t}, \bm{t}))$ distributed if  ${\bf K} = \Sigma(\bm{t}, \bm{t})^{-1}$ is Wishart distributed with degrees of freedom $\delta+p-1$ and scale matrix $\Psi(\bm{t}, \bm{t})^{-1}$.

The advantage of adopting the parameterization of \citet{dawid1981some} is two-fold: the parameter $\delta$ does not vary with the dimension of $\bm{t}$, and the resulting Inverse-Wishart distribution is consistent under marginalization. These properties are essential in showing that the IWP prior for $\Sigma(\cdot, \cdot)$ is well-defined when the number of grid points $p$ approaches infinity. This is summarized in Proposition \ref{prop}. Proposition \ref{prop} essentially states that $\mbox{IWP}(\delta, \Psi)$ is a well-defined (infinite-dimensional) probability measure and its finite-dimensional projection on grid $\bm{t}\times \bm{t}$ coincides with the Inverse-Wishart distribution $\mbox{IW}(\delta, \Psi(\bm{t},\bm{t}))$. The proof can be derived through following similar arguments as in the proof of Lemma 2 in the Appendix of \citet{zhu2012bayesian1}.

\begin{proposition}
  \label{prop}
  Let $\mathcal{T} \subseteq \mathbb{R}$ be a compact set and $\delta > 4$ be a positive integer. Suppose that $\Psi : \mathcal{T} \times \mathcal{T} \rightarrow \mathbb{R}$ is a symmetric and positive semi-definite mapping, i.e., any evaluation of $\Psi$ on a finite grid
  $\bm{t}  \times \bm{t} \subseteq \mathcal{T}  \times \mathcal{T} $ gives a symmetric and positive semi-definite matrix.
Then there exists a unique probability measure $\lambda$ on $(\mathbb{R}^{\mathcal{T}\times \mathcal{T}}, \mathcal{B}(\mathbb{R}^{\mathcal{T}\times \mathcal{T}}))$,
such that for any finite discretization $\bm{t}$,
$\lambda_{\bm{t}\times\bm{t}} = \mbox{IW}(\delta, \Psi(\bm{t},\bm{t}))$. We denote $\lambda=\mbox{IWP}(\delta, \Psi)$.
  \end{proposition}

In the prior distribution of $\Sigma \sim \mbox{IWP}(\delta, \Psi)$, a smaller value of $\delta$  leads to a less informative prior; the parameter $\Psi$ controls the prior covariance structure. To encourage a smoother estimation of the covariance, we set $\Psi(\cdot, \cdot) = \sigma_s^2 A(\cdot, \cdot)$, where $\sigma_s^2$ is a positive scaling parameter and $A(\cdot, \cdot)$ is a smooth correlation/covariance kernel which can be either stationary or nonstationary. The structure of $A(\cdot, \cdot)$ may be nonparametric or of parametric form with hyperparameters. In this paper, we use
the Mat\'{e}rn parameterization as an example of the stationary correlation structure, i.e., $ \Psi(t_i, t_j) =  \sigma_s^2 \,\mbox{Matern}_{cor}(|t_i - t_j|;  \rho,  \nu)$, where the Mat\'{e}rn correlation kernel $\mbox{Matern}_{cor}(d;  \rho,  \nu)$ is defined as
\begin{equation*}
\mbox{Matern}_{cor}(d; \rho, \nu) =  \frac{1}{\Gamma(\nu) 2^{\nu - 1}}
\left(\sqrt{2\nu}\frac{d}{\rho}  \right)^{\nu}  K_{\nu}\left( \sqrt{2\nu}\frac{d}{\rho}\right), \quad d\ge 0,\; \rho>0, \; \nu>0.
\end{equation*}
In the above expression, $d$ denotes the distance between two measurement points;
$\rho$ is the scale parameter; $\nu$ is the order; and $K_\nu(\cdot)$ is the modified Bessel function of the second kind. One of the important properties of 
the Mat\'{e}rn covariance kernel is its positive definiteness \citep{stein1999}. Both $\rho$ and $\nu$ can influence the smoothness of the signal estimates.
We choose $\nu > 2$ so that the resulting GP is $ \lfloor\nu - 1\rfloor$ times differentiable, which ensures the smoothness of the underlying data. It is also convenient to take $\nu$ to be an integer plus $0.5$, in which case
the Mat\'{e}rn correlation kernel has a closed-form expression.
For example, if $\nu = 2.5$, the correlation kernel $A(\cdot, \cdot)$ takes the form
\begin{eqnarray}
A(t_i, t_j)&=& \left( 1 + \frac{\sqrt{5} |t_i-t_j|}{\rho} + \frac{5(|t_i - t_j|)^2}{3\rho^2}\right)
\exp{\left(-\frac{\sqrt{5}|t_i - t_j|}{\rho}\right)}. \label{bayes:A}
\end{eqnarray}

Notice that the Mat\'{e}rn structure is sensitive to both $\rho$ and $\nu$, while the estimation of these parameters can be unstable. 
\citet{zhang2004inconsistent} has pointed out in a geostatistical setup that not all three parameters $(\sigma_s^2, \rho, \nu)$ in the Mat\'{e}rn class can be estimated consistently. To ensure a stable covariance estimation, we set $\rho$ and $\nu$ at fixed values obtained from empirical correlation estimation, by minimizing the mean square error between an empirical correlation estimate and the $\mbox{Matern}_{cor}(\rho, \nu)$ function. The hyperprior for $\sigma_s^2$ is set to be $\sigma_s^2 \sim \mbox{Gamma}(a_s, b_s)$, i.e., a Gamma distribution with shape parameter $a_s$ and inverse scale parameter $b_s$.

The above Bayesian hierarchical model is constructed based on infinite-dimensional GPs. However, practical posterior calculations can only be conducted in a finite manner. Since we assume latent GPs in our model, posterior inference can be performed similarly as in GP regression.  In particular, the latent processes $\{Z_i(\cdot)\}$ and the parameters $\mu(\cdot)$, $\Sigma(\cdot, \cdot)$ will be inferred on a finite grid, while the inference on non-grid points, if needed, can be obtained by posterior prediction.  
Therefore, we will represent the likelihood for model (\ref{bayes:eq2.1}) and the priors (\ref{eq:prior}) in multivariate forms through evaluating the functions on finite grids. Denoting $Y_i(\bm{t}_i)$ by $\bYi$, $Z_i(\bm{t}_i)$ by $\bZi$, model  (\ref{bayes:eq2.1}) implies that
\begin{eqnarray}
&\bYi \mid \bZi, \sigma_{\epsilon}^2  \sim
N(\bZi, \sigma_{\epsilon}^2 \mathbf{I}), \quad i=1, \cdots, n, & \label{eq:mvn1}\\
&\bZi \mid \mu(\bti), \Sigma(\bti, \bti)   \sim
N(\mu(\bti), \Sigma(\bti, \bti)), & \label{eq:mvn2}
\end{eqnarray}
where $\mathbf{I}$ is  a $p_i \times p_i$ identity matrix. Since the grids $\{\bm{t}_i; i= 1, 2, \cdots, n\}$ are not required to be common, we evaluate the GP and IWP prior distributions in (\ref{eq:prior}) on the pooled grid $\bm{t} = \cup_{i=1}^n \bm{t}_i$. Denote $\mu(\bm{t})$ by $\bm{\mu}$, $\mu_0(\bm{t})$ by $\bm{\mu}_0$,  $\Sigma(\bm{t}, \bm{t})$ by  $\bm{\Sigma}$ and $\Psi(\bm{t}, \bm{t})$ by $\bm{\Psi}$, then
\begin{equation}
\bm{\mu} \mid \bm{\Sigma} \sim N\left(\bm{\mu}_0, \frac{1}{c}\bm{\Sigma}\right), \quad
\bm{\Sigma} \sim \mbox{IW}(\delta, \bm{\Psi}).
\label{eq:priormvn1}
\end{equation}

The multivariate representations in (\ref{eq:mvn1})--(\ref{eq:priormvn1}) enable us to write the joint posterior distribution of
$(\{Z_i(\bm{t})\}, \bm{\mu}, \bm{\Sigma}, \sigma^2_{\epsilon}, \sigma_s^2 )$, based on which posterior sampling can be performed by MCMC; details are presented in Section \ref{Bayes:mcmc}. The posterior means of  $\{Z_i(\bm{t})\}$ will be treated as smoothed functional signals.

\subsection{Prior parameter setup}
\label{Bayes:prior}

The proposed Bayesian hierarchical model described in Section \ref{Bayes:assump} contains several hyper-parameters: $(c, \bm{\mu_0}, \nu, \rho,  a_{\epsilon}, b_{\epsilon},  a_s, b_s)$. Their values are determined using empirical methods. In particular,  we set $c = 1$, which implies that the functional mean has Gaussian prior with the same covariance kernel as the data. We set $\bm{\mu_0}$ to be the smoothed sample mean of $\{\bYi\}$ by existing methods. To facilitate stable covariance estimation, we take $\nu$ and $\rho$ as their empirical estimates $\widehat{\nu}$ and $\widehat{\rho}$ respectively.
The empirical estimates can be obtained by minimizing the mean square error between an empirical correlation estimate and a $\mbox{Matern}_{cor}(\bm{D}; \rho, \nu) $ correlation function with distance matrix $\bm{D}$, subject to $\rho > 0$ and $\nu \geq 2.5$.

The values of $(a_{\epsilon}, b_{\epsilon},  a_s, b_s)$
are determined using a heuristic empirical Bayes approach with empirical estimates of
 \{$\sigma^2_{\epsilon}$, $\sigma_s^2$\}, which is described as follows:

 \begin{itemize}
 \item The value of $\widehat{\sigma}^2_{\epsilon}$ can be easily obtained  using
 a differencing technique  \citep{von1941distribution}
\begin{equation}
\label{sigma_noise}
\widehat{\sigma}^2_{\epsilon}=
\frac{1}{2\sum_{i=1}^n(p_i-1)} \sum_{i=1}^n \sum_{j=1}^{p_i-1} (Y_i({t_{i(j+1)})} - Y_i(t_{ij}))^2.
\end{equation}
\item A moment estimator of $\sigma_s^2$ can be derived by
taking expectation (with respect to the prior distribution of $\bm{\Sigma}$) and  applying a trace operator to both sides of the equation
$\mbox{Cov}(Y(\bm{t})) = \bm{\Sigma} + \sigma_{\epsilon}^2\mathbf{I}$:
\begin{eqnarray}
\mbox{trace}(E\{\mbox{Cov}(Y(\bm{t}))\}) &= &\frac{\mbox{trace}(\bm{\Psi}) }{\delta-2} + \sigma_{\epsilon}^2 \mbox{trace}(\bm{I})
= \frac{\sigma_s^2 \mbox{trace}(\bm{A})}{\delta-2} + p\sigma^2_{\epsilon}, \nonumber \\
\widehat{\sigma}_s^2 & \approx &
\frac{\mbox{trace}(E\{\mbox{Cov}(Y(\bm{t}))\}) - p \widehat{\sigma^2_{\epsilon}}}{\mbox{trace}(\bm{A})/(\delta-2)}, \label{sigma_s2_hat}
\end{eqnarray}
where $p$ is the length of the pooled grid $\bm{t}$;
$\widehat{\sigma}^2_{\epsilon}$ is given by (\ref{sigma_noise}); and $E\{\mbox{Cov}(Y(\bm{t}))\}$ can be estimated by an empirical method, e.g.~the covariance estimate by PACE \citep{yao2005functional}.

\item The values of $(a_{\epsilon}, b_{\epsilon},  a_s, b_s)$ can be estimated by the method of moments with the prior distributions of $\sigma_{\epsilon}^2$, ${\sigma}_s^2$  and the empirical estimates in (\ref{sigma_noise}) and (\ref{sigma_s2_hat}).

\end{itemize}
\vspace{-0.1 in}

\section{Posterior inference with MCMC}
\label{Bayes:mcmc}

Based on the multivariate representations (\ref{eq:mvn1})--(\ref{eq:priormvn1}), we derive the joint posterior distribution and develop a MCMC algorithm for posterior inference in this section. Denote the observed data by $\bm{Y} = \{\bm{Y}_{\bm{t}_1}, \bm{Y}_{\bm{t}_2}, \cdots, \bm{Y}_{\bm{t}_n}\}$; denote the underlying GP evaluations on the observational grids by $\bm{Z} = \{\bm{Z}_{\bm{t}_1}, \bm{Z}_{\bm{t}_2}, \cdots, \bm{Z}_{\bm{t}_n}\}$; and denote the evaluations on the pooled grid by $\bm{\widetilde{Z}} = \{Z_1(\bm{t}), Z_2(\bm{t}), \cdots, Z_n(\bm{t})\}$. Recall that $\bm{t}=\cup_i \bm{t}_i $ is the pooled grid. Let $\bm{Z}^{\ast} =\bm{\widetilde{Z}}\setminus \bm{Z}$, i.e. 
$\bm{Z}^{\ast} = \{\bm{Z}_{\bm{t}_1^{\ast}} , \bm{Z}_{\bm{t}_2^{\ast}} , \cdots, \bm{Z}_{\bm{t}_n^{\ast}}\}$, where $\bm{Z}_{\bm{t}_i^{\ast}}  = Z_i(\bm{t}_i^{\ast})$ and $\bm{t}_i^{\ast}  = \bm{t} \setminus \bm{t}_i$ is the set of grid points for the $i$th trajectory with missing observations.   
The joint posterior density of all parameters can be written as 

\begin{eqnarray}
&& f(\bm{\widetilde{Z}}, \bm{\mu, \Sigma}, \sigma^2_{\epsilon}, \sigma_s^2 | \mathbf{Y}) \propto 
  f(\bm{Y}| \bm{\widetilde{Z}}, \sigma^2_{\epsilon})  
  f(\bm{\widetilde{Z}}| \bm{\mu, \Sigma}) f(\sigma^2_{\epsilon}) f(\bm{\mu|\Sigma})f(\mathbf{\Sigma}|
  \sigma_s^2) f(\sigma_s^2) \nonumber \\
&\propto &
  f(\bm{Y} | \bm{Z}, \sigma^2_{\epsilon})f(\bm{Z}^{\ast}  | \bm{Z}, \bm{\mu}, \bm{\Sigma})  
  f(\bm{Z} | \bm{\mu}, \bm{\Sigma}) f(\sigma^2_{\epsilon})
  f(\bm{\mu | \Sigma})
  f(\mathbf{\Sigma}|
  \sigma_s^2) f(\sigma_s^2). \label{eq:jointpost}
\end{eqnarray}  

Here we have factored the joint prior of $\bm{\widetilde{Z}}$ as  $f(\bm{\widetilde{Z}}|\bm{\mu, \Sigma}) = 
 f(\bm{Z}^{\ast} | \bm{Z}, \bm{\mu}, \bm{\Sigma})f(\bm{Z}| \bm{\mu}, \bm{\Sigma})$, which enables the updates of $\bm{Z}$ and $\bm{Z}^{\ast}$ by a Gibbs sampler. To design a MCMC algorithm of the Gibbs sampler, we need to derive conditional distributions for the latent variables $\bm{Z}$, $\bm{Z}^{\ast}$ and the model parameters $\bm{\mu}, \bm{\Sigma}, \sigma_s^2, \sigma^2_{\epsilon}$.  For brevity, we only present the conditional posterior distributions of $\bm{Z}$ and $\bm{Z}^{\ast}$ in Section \ref{sec:cond}. The derivation of the conditional posteriors for the remain parameters are trivial due to the conjugacy of their priors.  
  
\subsection{Conditional posteriors of $\bm{Z}$ and $\bm{Z}^{\ast}$}
\label{sec:cond} 

In case that all functional data are observed on a common grid, i.e.~$\{\bm{t_i} \equiv \bm{t}; i=1, 2, \cdots, n\}$,  then $\bm{Z}^*$ vanishes and the conditional posterior distribution of $\bZi$ can be derived from 
$$f(\bZi | \bYi, \bm{\mu, \Sigma}, \sigma_{\epsilon}^2) \propto  f(\bYi | \bZi, \bm{\mu, \Sigma}, \sigma_{\epsilon}^2) f(\bZi |\bm{\mu, \Sigma}),$$ 
which gives 
\begin{eqnarray}
\label{Zi_cgrid}
(\bZi | \bYi, \bm{\mu, \Sigma},  \sigma_{\epsilon}^2) &\sim& N(\widetilde{\bm{\mu_i}}, \widetilde{\bm{V_i}}), \\
\widetilde{\bm{V_i}} &=& \left((1/\sigma^2_{\epsilon} ) \bm{I} + \bm{\Sigma}^{-1}\right)^{-1}; \nonumber \\
\widetilde{\bm{\mu_i}} &=& \widetilde{\bm{V_i}} \left((1/\sigma^2_{\epsilon}) \bYi + \bm{\Sigma}^{-1}\mu(\bm{t_i})), \text{ where }  \bm{\Sigma} = \Sigma(\bm{t_i}, \bm{t_i} \right).
\nonumber
\end{eqnarray}

In case that functional data are collected on uncommon grids, we will first update $\bZis$ from 
\begin{eqnarray}
\label{Zis}
(\bm{Z}_{{\bf t}_i^{\ast}} | \bZi, \bm{\mu}, \bm{\Sigma}) &\sim & 
N(\bm{\mu_i}^{\ast}, \bm{V_i}^{\ast}), \\
\bm{\mu_{i}^{\ast}} & = & \mu(\bm{t_i^{\ast}}) +  \Sigma(\bm{t_i^{\ast}}, \bm{t_i})\Sigma(\bm{t_i}, \bm{t_i})^{-1}( \bm{Z_{t_i}} -\mu(\bm{t_i}) ) = \bm{B_i}\bZi - \bm{u_i},  \nonumber \\ 
&& \text{ where }\bm{B_i} =  \Sigma(\bm{t_i^{\ast}}, \bm{t_i})\Sigma(\bm{t_i}, \bm{t_i})^{-1}, \quad \bm{u_i} = \bm{B_i}\mu(\bm{t_i}) -   \mu(\bm{t_i^{\ast}}); \nonumber \\
\bm{V_{i}^{\ast}} & = & \Sigma(\bm{t_i^{\ast}}, \bm{t_i^{\ast}}) -  \Sigma(\bm{t_i^{\ast}}, \bm{t_i})\Sigma(\bm{t_i}, \bm{t_i})^{-1}\Sigma(\bm{t_i}, \bm{t_i^{\ast}}). \nonumber
\end{eqnarray}
Then update $\bZi$ from the conditional posterior distribution
\begin{equation}
\label{fZi}
f(\bZi | \bYi, \bm{\mu, \Sigma}, \sigma_{\epsilon}^2, \bZis) \propto  f(\bYi | \bZi, \bm{\mu, \Sigma}, \sigma_{\epsilon}^2) f(\bZi |\bm{\mu, \Sigma}) f(\bZis | \bZi, \bm{\mu, \Sigma}),
\end{equation}
which gives
\begin{eqnarray}
\label{Zi_sparse}
(\bZi | \bYi, \bZis, \bm{\mu}, \bm{\Sigma}, \sigma_{\epsilon}^2) &\sim & 
N(\widetilde{\bm{\mu_i}}, \widetilde{\bm{V_i}}), \\
\widetilde{\bm{V_i}} &=& \left( (1/\sigma^2_{\epsilon} ) \bm{I} + \Sigma(\bm{t_i}, \bm{t_i})^{-1} +  \bm{B_i}^T (\bm{V_{i}}^{\ast})^{-1}\bm{B_i} \right)^{-1} ; \nonumber \\
\widetilde{\bm{\mu_i}} &=& \widetilde{\bm{V_i}} \left( (1/\sigma^2_{\epsilon}) \bYi + \Sigma(\bm{t_i}, \bm{t_i})^{-1}\mu(\bm{t_i}) +   \bm{B_i}^T (\bm{V_{i}}^{\ast})^{-1} (\bm{u_i} + \bZis) \right). \nonumber
\end{eqnarray}

When $\bm{t^{\ast}}$ and $\bm{t}$ are both dense, the conditional variance $\bm{V_i^{\ast}}$ in (\ref{Zis}) can be very close to a zero matrix, which could cause numerical instability when inverting $\bm{V_i^{\ast}}$. In this case, we suggest using the conditional mean in (\ref{Zis}) as a sample for $\bZis$ in the MCMC algorithm and removing the conditional prior $ f(\bZis | \bZi, \bm{\mu, \Sigma})$ from (\ref{fZi}) when updating $\bZi$ (note that now (\ref{Zi_sparse})  collapses to (\ref{Zi_cgrid})).

\subsection{MCMC algorithm}
\label{sec:mcmc}

Based on the joint posterior distributions derived from (\ref{eq:jointpost}), we design a MCMC algorithm for posterior sampling as follows:
\vspace{-0.1 in}
\begin{itemize}[leftmargin=0.0 in]
\item[] Step 0: Set initial values. Set
$(\bm{\mu}, \sigma_{\epsilon}^2)$ to be the empirical estimates, $\bm{Z}$ to be the raw data $\bm{Y}$, and $\bm{\Sigma}$ to be an identity matrix. The prior parameters $(c, \bm{\mu_0}, \nu, \rho,  a_{\epsilon}, b_{\epsilon},  a_s, b_s)$ are set as described in Section \ref{Bayes:prior}. 

\item[] Step 1: Conditional on $\bm{Y}$ and current values of $(\bm{\mu},\bm{\Sigma})$, update $\bm{Z}^*$ and $\bm{Z}$. 
In the general case where all data are observed on uncommon grids, update  $\bm{Z}^*$ and  $\bm{Z}$  from (\ref{Zis}) and (\ref{Zi_sparse})  alternatively. In the case of common grids, update $\bm{\widetilde Z}$ (identical to $\bm{Z}$) from  (\ref{Zi_cgrid}).

\item[] Step 2: Conditional on $\bm{Y}$ and current values of $\bm{Z}$, 
update the noise variance $\sigma^2_{\epsilon}$ by
  $$(\sigma^2_{\epsilon} | \bm{Y}, \bm{Z}) \sim \mbox{Inverse-Gamma}\left(a_{\epsilon}+
    \frac{\sum_{i=1}^n p_i}{2},\quad  b_{\epsilon} + \frac{1}{2}
    \sum_{i=1}^n\left[ (\bm{Y_{t_i}} -  \bm{Z_{t_i}})^T (\bm{Y_{t_i}} -  \bm{Z_{t_i}}) \right]\right).$$

\item[] Step 3: Conditional on current values of $\widetilde{\bm{Z}}= \bm{Z}\cup \bm{Z}^*$ and $\bm{\Sigma}$, update $\bm{\mu}$ from 
 $$(\bm{\mu}|\widetilde{\bm{Z}}, \bm{\Sigma})  \sim N\left(\frac{1}{n+c}
    \left(\sum_{i=1}^n Z_i(\bm{t})  + c \bm{\mu_0} \right),\quad \frac{1}{n+c}
    \mathbf{\Sigma} \right).$$
    
\item[] Step 4: Conditional on current values of $\widetilde{\bm{Z}}= \bm{Z}\cup \bm{Z}^*$ and $\bm{\mu}$, update $\bm{\Sigma}$ from
  $$(\mathbf{\Sigma}|\widetilde{\bm{Z}}, \bm{\mu}) \sim IW
    \left(n+\delta+1, \bm{Q} \right), \quad     
  \bm{Q} = (\widetilde{\bm{Z}}-\bm{\mu} \bm{J})
  (\widetilde{\bm{Z}}-\bm{\mu} \bm{J})^T +
   c(\bm{\mu} - \bm{\mu_0}) (\bm{\mu} - \bm{\mu_0})^T + \sigma^2_s \bm{A},
  $$
   where $\widetilde{\bm{Z}}$ denotes a matrix with columns $\{Z_i(\bm{t})\}$; $\bm{J} = (1, \cdots, 1)$  is a vector of ones with length $p$;
and matrix $\bm{A}$ is given by equation (\ref{bayes:A}).

\item[] Step 5:  Given current values of $\mathbf{\Sigma}$, update $\sigma_s^2$  from 
$$  (\sigma_s^2| \mathbf{\Sigma})
  \sim \mbox{Gamma}\left(a_s + \frac{(\delta+p-1) p}{2}, \quad b_s +
  \frac{1}{2}\,\mbox{trace}(\bm{A}\bm{\Sigma}^{-1})\right). $$
\end{itemize}
\vspace{-0.1 in}

\noindent Repeat Steps $1\sim5$ for a large number of iterations until convergence is achieved.  
  
For the simulations and real case studies in Sections \ref{Bayes:sim} and \ref{Bayes:real}, we ran  $10,000$ MCMC iterations after a burn-in period of $2,000$ iterations. The convergence of the MCMC chains were diagnosed by the Gelman and Rubin diagnostic method \citep{gelman1992inference, Simo2014}.

\section{Simulation studies}
\label{Bayes:sim}

Two simulation studies were performed for functional data with stationary and nonstationary covariance, respectively. Within each study, we analyzed functional observations on both common and uncommon grids.
\FloatBarrier
\subsection{Simulation 1: functional data with stationary covariance}
\label{Bayes:sim:gp}

Two types of functional data with stationary covariance were generated with common and uncommon grids. The underlying smooth functional data were generated from a GP with mean $\mu(t) = 3 \sin(4t)$ and covariance kernel $\Sigma(s, t) = 5\,\mbox{Matern}_{cor}(|s - t|;\rho,  \nu)$, where $\rho = 0.5$, $\nu = 3.5$, and $s, t\in[0, \pi/2]$. Each dataset contained $n=50$ observations. The common grid contained $p=80$ equally spaced grid points on $[0, \pi/2]$.  Noise terms $\{\epsilon_{ij}; i = 1, \cdots, n, j = 1, \cdots, p\}$ were independently generated from $N(0,  \sigma_{\epsilon}^2)$ with $\sigma_{\epsilon} = \sqrt{5} / 2$, and then added to the smooth functional data, resulting the final dataset with SNR equal to $2$. In the uncommon grid case, we randomly retained data on $60\%$ of the common grid points, which resulted in moderately sparse functional data on uncommon grids.  Two sample curves for the common and uncommon grid cases are shown in Figure \ref{fig:1}(a, b), where the gray lines/line-segments denote the raw data.

Based on the data generated above, we applied the MCMC algorithm in Section \ref{sec:mcmc} to obtain posterior samples. Taking the common grid case as an example, we set $\delta = 5$, $\widehat{\rho} = 0.503$,  $\widehat{\nu} = 3.459$, $\widehat{\sigma}^2_{\epsilon} = 1.239$, and $\widehat{\sigma}_s^2 = 10.750$ using empirical estimation methods. With $\widehat{\sigma}^2_{\epsilon}$ and $\widehat{\sigma}_s^2$, we obtained values for the hyperprior parameters 
\{$a_{\epsilon} = 0.807$, $b_{\epsilon} = 1$, $a_s=214.990$, $b_{s} =20$\}.
The posterior means of the parameters were calculated by averaging $10, 000$ posterior samples. The parameter $\sigma_{\epsilon}^2$, with true value 1.250, had posterior mean $1.244$ and 95\% credible interval $[1.187, 1.302]$. The parameter $\sigma_s^2$, with unidentifiable true value, had posterior mean $73.443$ and 95\% credible interval $[69.700, 77.620]$. 

In Figure \ref{fig:1}(a, b), we display the smoothed curves (black solid lines), together with the raw data (gray lines/line segments), the 95\% pointwise credible intervals (black dash-dot lines), and the underlying true smooth curves (blue dots) for the common (frame (a)) and uncommon grid (frame (b)) cases.  The heat maps of our Bayesian correlation estimates are shown in Figure \ref{fig:1}(c, d) for common (frame (c)) and uncommon (frame (d)) grid cases. In contrast, we also plotted the heat map of the sample correlation in the common grid case in Figure \ref{fig:1}(e), and the heat map of the underlying true correlation in Figure \ref{fig:1}(f).  These plots show that our method can recover the true smooth signals accurately for both common and uncommon grid cases, while providing a much smoother correlation estimate than the sample estimate in Figure \ref{fig:1}(e). Even though around $40\%$ of the grid points are missing in each trajectory, the results of smoothing and mean-covariance estimation as in Figure \ref{fig:1}(a, d) are almost as good as the common grid results as in Figure \ref{fig:1}(b, c).

\begin{figure}[htb]
\begin{center}
\includegraphics[width= 0.93\linewidth]{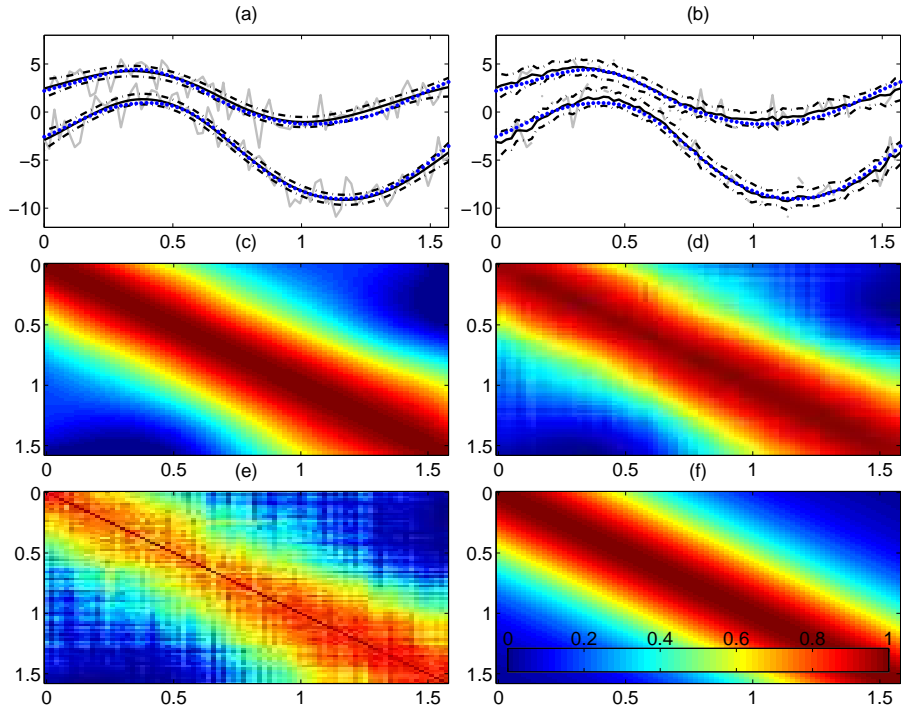}
\caption{Simulation 1: (a) Two sample curves for functional observations in the common grid case (gray lines), superimposed by the Bayesian smoothed curves (black solid lines), along with the 95\% pointwise credible intervals (black dashed lines) and the true signals (blue dots); (b) plots (same as in (a)) for the uncommon grid case; (c) heat map of our Bayesian correlation estimate in the common grid case; (d) heat map (same as in (c)) in the uncommon grid case; (e) heat map of the sample correlation in the common grid case; (f) heat map of the true underlying (Mart\'ern) correlation.
}
\label{fig:1}
\end{center}
\end{figure}


To further assess the performance of the proposed method, we compared the results with four alternative smoothing methods: {\it best possible least square} (BLS) estimation, smoothing spline with cubic splines applied to each individual curve (Spline), kernel smoothing with local polynomials applied to each individual curve (Kernel), and PACE.  The BLS method served as an ``oracle" model, in which the smoothed curves were estimated by the conditional mean of the latent GPs, while assuming the true mean and covariance of the latent processes were known. In the common grid case, the conditional mean and covariance for each signal can be written as
\begin{eqnarray*}
E\{\bm{Z}_i | \bm{Y}_i, \bm{\mu}, \bm{\Sigma}\} &=& \bm{\mu} +
\bm{\Sigma} (\bm{\Sigma} + \sigma_{\epsilon}^2 \bm{I})^{-1} (\bm{Y}_i - \bm{\mu}),\\
\mbox{Cov}(\bm{Z}_i | \bm{Y}_i, \bm{\mu}, \bm{\Sigma}) &=&
\bm{\Sigma} -  \bm{\Sigma}(\bm{\Sigma} + \sigma_{\epsilon}^2 \bm{I})^{-1}\bm{\Sigma},
\end{eqnarray*}
which can be easily derived from the joint Gaussian distribution of $\bm{ Y}_i$ and  $\bm{Z}_i$.
We obtained the Spline and Kernel estimates using R  \citep{R} functions \verb#smooth.spline# and \verb#locpoly# from the \verb#KernSmooth# library. The smoothing parameter in the former function was determined by generalized cross-validation (GCV) and the smoothing parameter in the latter function was selected by a direct plug-in approach using the \verb#dpill# function. We chose \verb#degree = 1# in the \verb#locpoly# function (equivalent to local linear smoothing). For the Spline and Kernel methods, we estimated the mean curve by averaging individually smoothed curves. The PACE method was implemented with a Matlab package developed by \citet{yao2005functional} (\verb#http://www.stat.ucdavis.edu/PACE/#).

\begin{table}[ht]
\caption{Simulation results in the common grid case: mean RIMSEs and corresponding standard errors (in parentheses) of $Z_i(\bm{t})$ and $\mu(\bm{t})$, produced by Spline, Kernel,  PACE, our proposed methods (Bayesian), and BLS. The mean RIMSEs of the two best estimates are bold.}
\vspace{-0.1 in}
\begin{center} {
\begin{tabular}{lllllll}
\hline
  &  &  Spline & Kernel & PACE & Bayesian & BLS\\
\hline
 Stationary & $Z_i(\bm{t})$  & 0.4272 & 0.4591 & 0.4072 & \textbf{0.3783}  & \textbf{0.3649}   \\
 & &(0.0021) & (0.0016) & (0.0017) & (0.0014)  &(0.0014)\\
 
& $\mu(\bm{t})$ & 0.3819 & 0.3923	&0.3902	&\textbf{0.3781}	&\textbf{0.3643}     \\
 & &(0.0144) &(0.0150) &(0.0150) & (0.0144) &(0.0146)\\
\hline 
Nonstationary&$Z_i(\bm{t})$ &  0.5176 &  0.5453 & 0.4614 & \textbf{0.4599} & \textbf{0.4137}\\
 & &(0.0021) &(0.0016)  & (0.0023) & (0.0021) &(0.0016)\\

&$\mu(\bm{t})$&  0.5616	& 0.5738 	&  0.5802	& \textbf{0.5539}	&\textbf{0.5418} \\
 & &(0.0263) &(0.0264)	&(0.0261)   & (0.0264) &(0.0267)\\
\hline
\end{tabular}}
\end{center}
\label{mse-curve}
\end{table}

\begin{table}[ht]
\caption{Simulation results in the common grid case: mean RIMSEs and corresponding standard errors (in parentheses) for the estimates of $\Sigma(\bm{t},\bm{t})$ and $Cor(\bm{t}, \bm{t})$,  produced by  sample estimation with raw curves (Sample), PACE method, the proposed method (Bayesian), and sample estimation with Bayesian-smoothed curves (SE-Bayesian). The mean RIMSEs of the two best estimates are bold. }
\begin{center}
\begin{tabular}{llcccc}
\hline
 & Data  & Sample & PACE & Bayesian & SE-Bayesian \\
\hline
Stationary & $\Sigma(\bm{t}, \bm{t})$  & 1.5479   &  \textbf{1.2683}  & 2.2758 & \textbf{1.2523}\\
 && (0.0416)  & (0.0464) & (0.1803)&  (0.0461) \\
 
  & $Cor(\bm{t}, \bm{t})$& 0.2616   &  \textbf{0.1493} & \textbf{0.1410} & 0.1506\\
&& (0.0052)  & (0.0065) & (0.0066)&  (0.0068) \\

\hline

Nonstationary &$\Sigma(\bm{t}, \bm{t})$ &  2.5908 &  \textbf{2.3143}  & 2.7586  & \textbf{2.3456}\\
&  & (0.0758) & (0.0787) & (0.0951) & (0.0790)\\

&$Cor(\bm{t}, \bm{t})$ &  0.2625 &  \textbf{0.1661}  & \textbf{0.1695} & 0.1716\\
&  & (0.0043) & (0.0054) & (0.0056) & (0.0056)\\

\hline
\end{tabular}
\end{center}
\label{mse-cov}
\end{table}%

\begin{table}[ht]
\caption{Coverage probabilities for the 95\% pointwise credible intervals of $Z_i(\bm{t})$, $\mu(\bm{t})$,
and $\Sigma(\bm{t}, \bm{t})$.}
\begin{center}
\begin{tabular}{llll}
\hline
 Data  &  $Z_i(\bm{t})$ & $\mu(\bm{t})$ & $\Sigma(\bm{t}, \bm{t})$\\
\hline
    Stationary (common-grid)  &   0.9373 &  1.0000   & 0.9983   \\
    Stationary (sparse)  &  0.9685 & 0.9875    & 0.9192   \\
    Nonstationary (common-grid)  &0.9320&  0.8750  & 0.9730 \\
    Nonstationary (sparse)   & 0.9345  & 0.8625 & 0.9788 \\
\hline
\end{tabular}
\end{center}
\label{cov-prob}
\end{table}%

To quantitatively measure the goodness of smoothing and mean-covariance estimation, we repeated the above simulation 100 times for the common grid case. We calculated the root integrated mean square errors (RIMSEs), defined by $(\int_0^{\pi/2} (\widehat{Z}(t) - Z(t) )^2 dt)^{1/2}$ for signal estimates and by $(\int_0^{\pi/2}\int_0^{\pi/2} (\widehat{\Sigma}(s,t) - \Sigma(s,t) )^2 ds dt)^{1/2}$ for surface estimates, where the integrations were approximated by the trapezoidal rule. We compared the mean RIMSEs of our Bayesian estimates with the ones from alternative methods in the ``Stationary" sections in Tables 1 and 2.

We expect the BLS method provides the smallest mean RIMSEs and standard errors of $Z_i(\bm{t})$ and $\mu(\bm{t})$ as shown in Table \ref{mse-curve},  because BLS is an ``oracle" method using the true underlying distribution. The resulting mean RIMSEs and the standard errors from BLS should be treated as lower bounds that any statistical methods could achieve. We can also observe that our Bayesian approach achieves evident improvement on smoothing than the Spline/Kernel/PACE methods (with mean RIMSE  $0.3783$ vs.~$0.4272/0.4591/0.4072$). The mean curve estimate of the Bayesian method is also slightly better than the Spline/PACE/PACE methods ($0.3781$ vs.~$0.3819/0.3923/0.3902$).

In Table \ref{mse-cov}, we compared the mean RIMSEs from the Bayesian method with the ones from PACE, sample estimation with raw data (Sample), and sample estimation with Bayesian-smoothed curves (SE-Bayesian). We can observe that the SE-Bayesian estimate of covariance gives the lowest mean RIMSE, which is much smaller than the RIMSE of the direct Bayesian estimate ($1.2523$ vs.~$2.2758$). Here the relatively large mean RIMSE of the direct estimate from the Bayesian method may due to the non-identifiability between $\sigma_s^2$ and $\Psi$ (the scale covariance kernel in the IWP prior of $\Sigma$). However, the direct Bayesian estimate of the correlation surface has the smallest mean RIMSE, even slightly better than the PACE estimate (0.1410 vs.~0.1493).  In addition, the SE-Bayesian covariance estimate is comparable with the PACE estimate ($1.2523$ vs.~$1.2683$). This is not a surprise to us because both the proposed Bayesian method and PACE estimate a smooth covariance kernel from noisy (sparse) functional data.

In addition, we calculated the coverage probabilities for the Bayesian 95\% credible intervals of $Z_i(\bm{t})$, $\mu(\bm{t})$, and $\Sigma(\bm{t}, \bm{t})$, as shown in Table \ref{cov-prob}. We can see that the coverage probabilities are above 90\% for the simulation study with stationary covariance. From simulation studies, we find that the coverage probability for  $\mu(\bm{t})$ varies along the value of $c$ in (\ref{eq:prior}), which controls the magnitude of the prior variance of $\mu(\cdot)$.

\subsection{Simulation 2: functional data with nonstationary covariance}

\label{Bayes:sim:nongp}
We generated functional data with nonstationary covariance by imposing a nonlinear transformation on the true underlying GPs in Section \ref{Bayes:sim:gp}. Given the stationary GP $\widetilde{X}_i(t)$ in Section \ref{Bayes:sim:gp}, a nonstationary GP can be obtained by taking $X_i(t) = h(t)\widetilde{X}_i(\xi(t))$ with $h(t) = t + 1/2$ and  $\xi(t) = (t)^{2/3}$. As a result, $X_i(t)$ is a GP with mean $\mu(t)=3h(t)\sin(4 \xi(t))$ and covariance $\Sigma(s, t) = 5 h(s)h(t)\mbox{Matern}_{cor}(|\xi(s)-\xi(t)|;   \rho=0.5, \nu=3.5)$.
Similarly as in Simulation 1, $50$ nonstationary functional trajectories were generated on the same common grid, and noises were added to the true smoothed curves. Functional data in uncommon grid (sparse) case were generated in the same way as in Simulation \ref{Bayes:sim:gp}.

\begin{figure}[ht]
\begin{center}
\includegraphics[width= 0.93\linewidth]{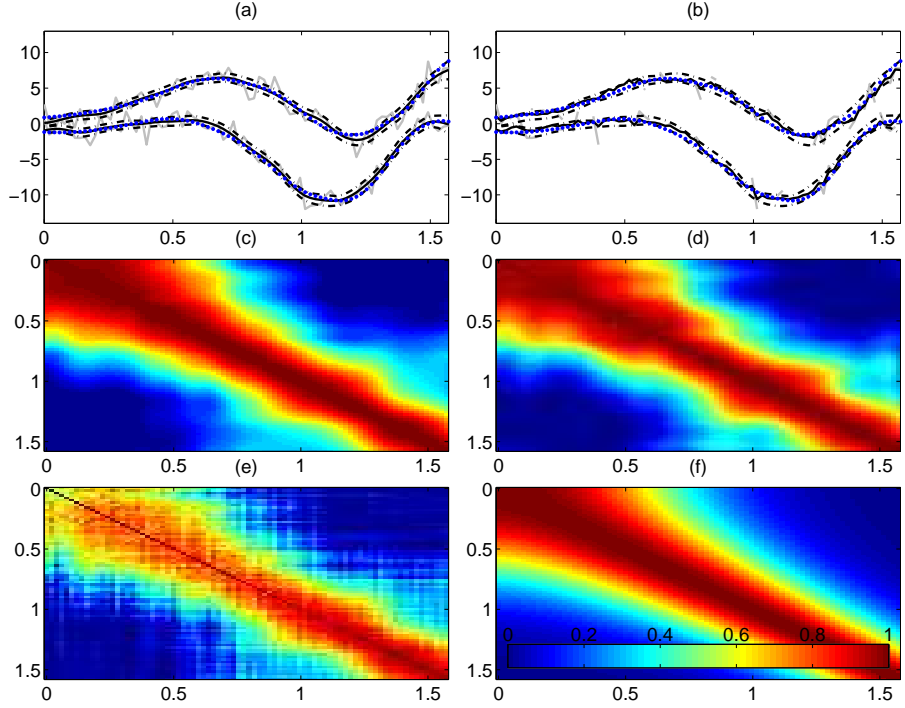}
\caption{Simulation 2: (a) two sample curves in the common grid case (gray lines), superimposed by the Bayesian estimates (black solid lines), along with the 95\% pointwise credible intervals (black dash-dot lines) and the true signals (blue dots); (b) curve plot (same as (a)) for the uncommon grid case; (c) heat map of our Bayesian correlation estimate in the common grid case; (d) heat map (same as (c)) in the uncommon grid case; (e) heat map of the sample correlation estimate (common grid case); (f) heat map of the true underlying (nonstationary) correlation surface. }
\label{fig:3}
\end{center}
\end{figure}


Two sample curves for the common and uncommon grid cases are plotted in Figure \ref{fig:3}(a, b).
We followed the same method of determining the hyperparameters and calculating the posterior means of the mean curve and covariance surface as in Section \ref{Bayes:sim:gp}. Particularly, for nonstationary data, it is desirable to choose $\Psi$ with a more flexible nonstationary structure. In this case study, we let $\Psi(\cdot,\cdot)=\sigma_s^2 A(\cdot, \cdot)$, and set $A(\cdot, \cdot)$ to be the covariance estimate obtained from PACE. Note that $A(\cdot, \cdot)$ has to be symmetric and positive definite.

Examples of the Bayesian estimates are shown in Figure \ref{fig:3}(a, b, c and d). We also plotted the heat map of the sample correlation in the common grid case in Figure \ref{fig:3}(e) and the heat map of the true underlying correlation in Figure \ref{fig:3}(f). The mean RIMSEs (from $100$ repeated simulations of the common grid case) for $\{Z_i(\bm{t})\}$, $\mu(\bm{t})$, and  $\Sigma(\bm{t}, \bm{t})$ are displayed in the ``Nonstationary" sections of Tables 1 and 2.  In addition, the coverage probabilities of the 95\% Bayesian credible intervals for $\{Z_i(\bm{t})\}$, $\mu(\bm{t})$, and $\Sigma(\bm{t}, \bm{t})$ are displayed in the ``Nonstationary" section of Table \ref{cov-prob}. The nonstationary results in Tables 1, 2 and 3 show similar patterns as in the stationary case studies. In particular, the proposed Bayesian approach produces the best smooth signal and mean estimates with the lowest mean RIMSEs.

\section{Real case studies}
\label{Bayes:real}

We use two real datasets with different levels of noise to show that our proposed Bayesian method produces accurate smooth estimates for the functional signals and mean-covariance. 

\subsection{Spectroscopy data with low levels of noise}
\label{spec:real}

In this application, we analyzed the spectroscopy data that were produced in a cervical pre-cancer study. The goal of the study was to diagnose early-stage cervical cancer with spectroscopy measurements \citep{yamal2012accuracy}. This data contain $n=462$ spectroscopy measurements produced by a multi-spectral digital colposcopy (MDC) device \citep{buys2012optical}. When taking the measurements, an operator puts a probe in contact with the cervical tissues; the device then ejects a beam of light through the probe onto the tissues and records the spectrum intensities of the reflected light through a white-light filter in the device. Each measurement contains intensity values (\verb#log10# transformed) ranging over emission wavelengths from 410 to 700 nanometer (nm). The spectroscopy data are collected on dense emission wavelengths (a case of common grid). We took intensity values at 1/3 of the equally spaced emission wavelengths as our observations (training data), and treated the values at the remaining 2/3 wavelengths as the validation data for prediction purposes. Figure \ref{fig:0}(a) shows three raw curves on their original scale (before the \verb#log10# transformation).

We then applied our Bayesian method on the training data (\verb#log10#  transformed), with $\delta = 5$ and hyperpriors set by the heuristic empirical Bayes method in Section \ref{Bayes:prior}. The values of the prior parameters were taken as: $ a_{\epsilon} = 1117.759$, $b_{\epsilon} = 1$, $a_s= 0.905$, $b_{s} =5$, $\widehat{\rho} = 369.716$, $\nu = 3.190$. The Bayesian estimates for $\sigma_{\epsilon}^2$ and $\sigma_s^2$ were $9.339 \times 10 ^{-5}$ with 95\% credible interval $(9.199\times 10 ^{-5}, 9.488\times 10 ^{-5})$, and $196.655$ with 95\% confidence interval $(173.679, 220.693)$, respectively. 

\begin{figure}[tb]
\begin{center}
\includegraphics[width=0.93\textwidth]{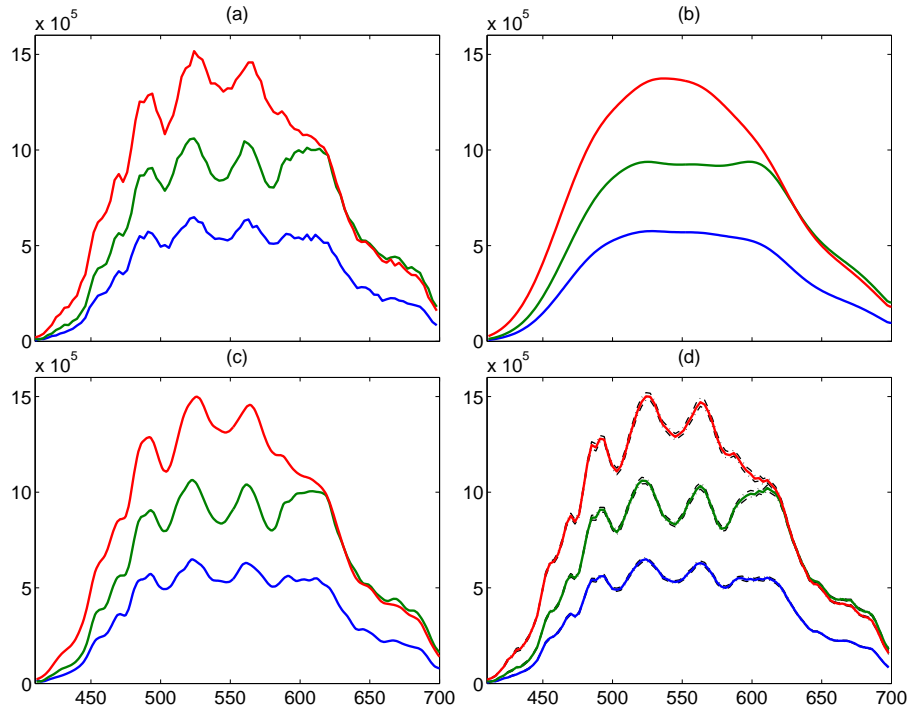}
\caption{Case study on spectroscopy data: (a) three raw spectroscopy curves;
(b) smooth estimates by the Kernel method; (c) smooth estimates by the Spline method; (d) smooth estimates by our Bayesian method along with 95\% pointwise credible intervals (black dashed curve).}
\label{fig:0}
\end{center}
\end{figure}

 We plotted the smoothed curves by the Kernel method in Figure \ref{fig:0}(b), the ones by the Spline method in Figure \ref{fig:0}(c), and the ones
by our Bayesian method in Figure \ref{fig:0}(d) together with 95\% pointwise credible intervals. One can observe that all local features in the raw data are completely wiped out by the kernel estimates in Figure \ref{fig:0}(b), which may due to the over-estimation of the smoothing parameter. In contrast, both the Spline and the Bayesian methods successfully preserved these features. However, because of the inherent nature of independently smoothing in the Spline method, the amount of smoothness varies across curves. For example, in Figure \ref{fig:0}(c), the red and green curves  appear to be slightly smoother than the blue curve, while our Bayesian method produces signal estimates with about the same amount of smoothness. This real data analysis demonstrates that our proposed Bayesian method is capable of preserving shared features by borrowing strength across all observations. 

\begin{figure}[h]
\begin{center}
\includegraphics[width=0.93\textwidth]{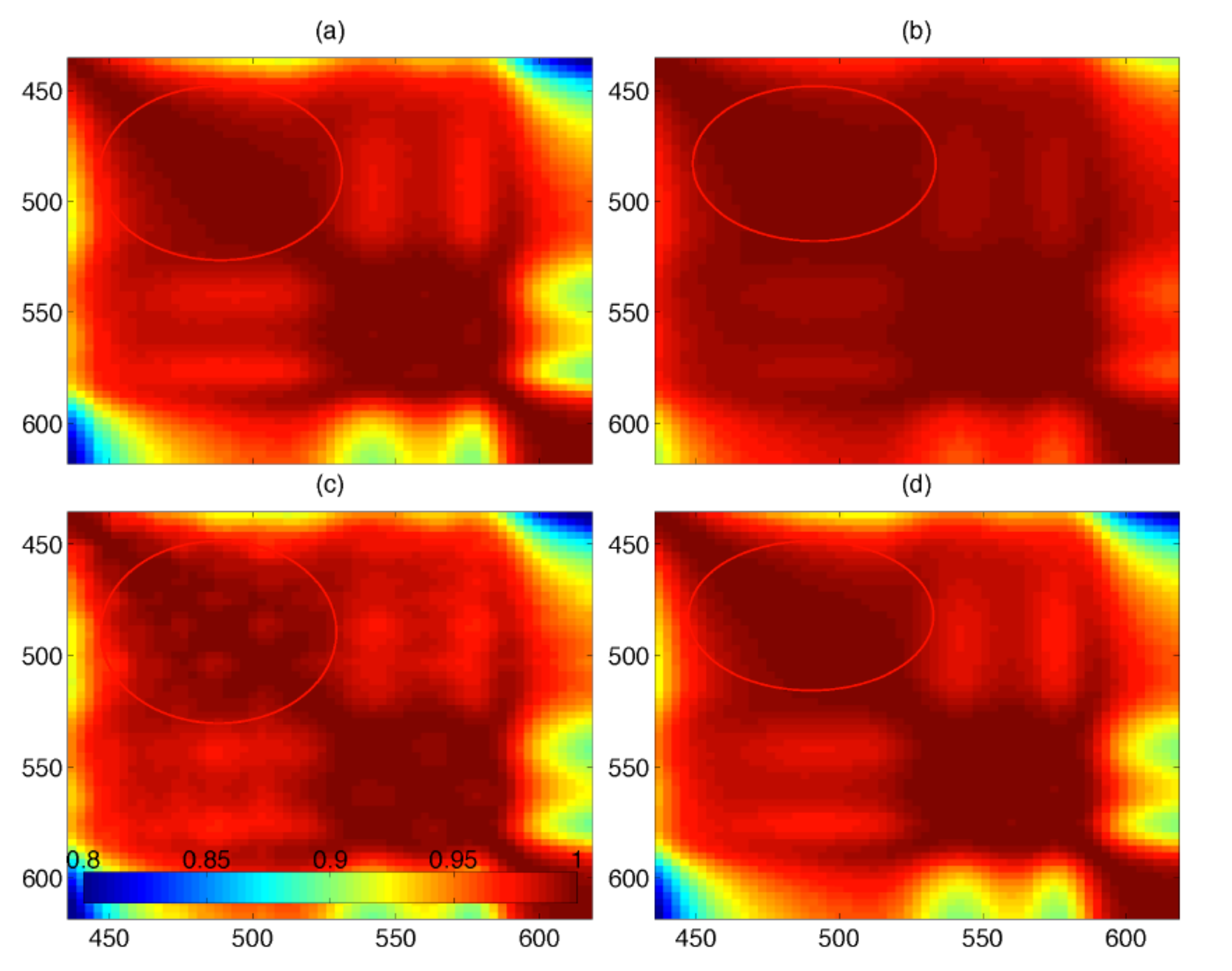}
\caption{Heat maps of the correlation estimates by the sample estimation with raw data in (a), Bayesian method in (b), PACE in (c), and sample estimation with Bayesian smoothed data in (d). }
\label{fig:5}
\end{center}
\end{figure} 

Due to low levels of noise, the influence of smoothing on the covariance estimation with this spectroscopy data is almost negligible. In such a situation one would expect a good correlation estimate to have a similar structure as the sample estimate.  Figure \ref{fig:5} shows the heat maps of the correlation estimates given by sample estimate with raw curves in frame (a), Bayesian method in frame (b), PACE in frame (c) and sample estimate with Bayesian smoothed curves in frame (d). The Bayesian estimate (Figure \ref{fig:5}(b)) and the sample estimate with Bayesian smoothed curves (Figure \ref{fig:5}(d)) are very close to the sample estimate based on the raw data (Figure \ref{fig:5}(a)). Although the PACE estimate in Figure \ref{fig:5}(c) shows a similar main structure, it appears to have different details around emission wavelength $500$nm (marked by red circles in Figure \ref{fig:5}), from the other three estimates. 

We used the remaining $2/3$ validation data to assess the prediction performance of the Bayesian method, compared with three alternative methods -- Spline, Kernel and PACE.  For the Spline method, the values on the validation grid points could be easily predicted with the estimated splines coefficients. For the Kernel/PACE/Bayesian methods, predicted values at the validation points were obtained using linear interpolation. Due to low noise levels in the spectroscopy data, we simply treated the observed values as true data and calculated the RMSEs as in Table \ref{mse-spec}. Here, the Bayesian method achieves the smallest RMSE, demonstrating the benefit of simultaneous smoothing on prediction. 

\begin{table}[h]
\caption{RMSEs for predictions on the validation grids.}
\begin{center}
\begin{tabular}{cccc}
\hline
   Spline  & Kernel & PACE & Bayesian \\ 
\hline
    0.020 &  0.039 &  0.025&  0.017 \\
\hline
\end{tabular}
\end{center}
\label{mse-spec}
\end{table}

\subsection{Metabolic data with high levels of noise}
We also studied a metabolic dataset from a study of obese children. The data were collected in the Children's Nutrition Research Center (CNRC) at the Baylor College of Medicine. We considered a subset of the data that contained energy expenditure (EE) measurements of 44 obese children. These measurements were collected every minute over a 100-minute period during their sleeping time. Therefore, each trajectory can be treated as the evaluation of an energy function over time. These raw curves appear to fluctuate frequently, implying potentially high levels of noise; three sample raw trajectories are plotted with the color gray in Figure \ref{fig:6}(a).

\begin{figure}[tb]
\begin{center}
\includegraphics[width=0.93\textwidth, height = 0.35\textheight]{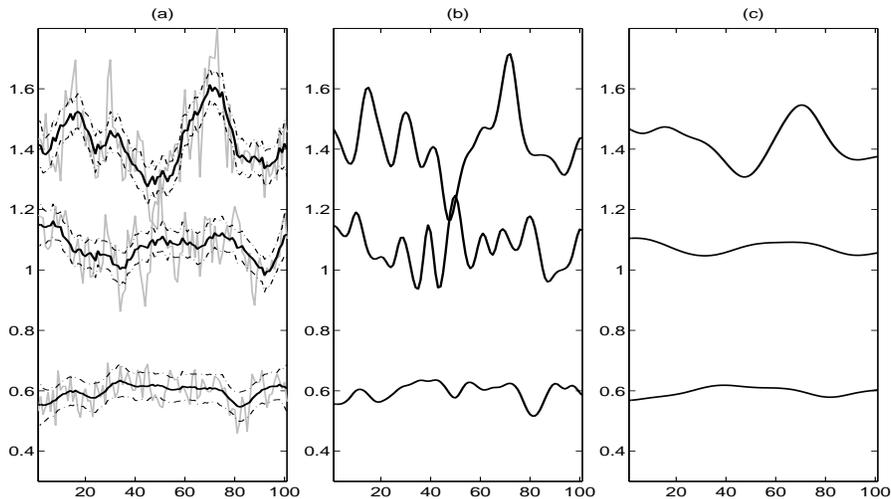}
\caption{ (a): Bayesian signal estimates (black solid) with 95\% credible intervals (black dashed) and raw observations (gray); (b): signal estimates by the Spline method;
(c): signal estimates by the Kernel method.}
\label{fig:6}
\end{center}
\end{figure} 

We applied the proposed Bayesian method to this dataset, using the common grid model  ($n < p$). We set $\delta = 5$ and set $A(\cdot, \cdot)$ as a Mat\'{e}rn correlation kernel, which led to signal estimates with similar level of variations (the black solid curves in Figure \ref{fig:6}(a)). In comparison with the Bayesian method, the Spline method produces excessive local variations in all three curves (Figure \ref{fig:6}(b)), and the Kernel smoothing method shows clear over-smoothing for all three curves (Figure \ref{fig:6}(c)).


\section{Discussion}
\label{Bayes:disc}

We have proposed a model-based  nonparametric Bayesian method to smooth all functional observations simultaneously and estimate the mean and covariance functions based on GPs. Specifically, we adopted a hierarchical framework by assuming a GP prior for the mean function and an IWP prior for the covariance function. The proposed method facilitats borrowing strength across all functional observations in simultaneous smoothing and an automatic mean-covariance estimation in the posterior inference.


Simulation studies show that our method produces smoothed signal estimates that are the closest to the ``oracle'' BLS estimates among all methods compared, and the resulting smoothed data lead to improved covariance estimation. In the real case study with spectroscopy data, our method demonstrates the most accurate prediction of function values at the validation wavelengths among all methods compared. The real case study with metabolic data shows that the smoothed signal estimates by our approach retain consistent patterns, while the alternative methods lost systematic trends common across all curves.

In our approach, posterior inference is performed by evaluating GPs on finite grids. 
To evaluate them on a new grid in posterior sampling in the most convenient way, one needs to perform posterior predictive sampling based on the conditional Gaussian distribution for smoothed data and to conduct a two-dimensional interpolation for the covariance surface while restricting the interpolated surface to be symmetric and positive definite.
 When dealing with functional data on uncommon grids, our approach involves evaluating mean and covariance functions on the pooled grid, which can be extremely dense and cause numerical problems. For example, the covariance matrices on dense grids are likely to be singular, thus inverting these matrices in (\ref{Zi_cgrid}), (\ref{Zis}), (\ref{Zi_sparse}) would be problematic. In such a case, sampling from a multivariate normal distribution with a nearly singular covariance matrix and sampling from an Inverse-Wishart distribution (in the MCMC steps) with a nearly singular scale matrix might also be challenging. These are common numerical issues encountered in GP models due to dense grids. 

Many solutions have been proposed to deal with the intensive computational burden and numerical issues caused by dense grids. The essential idea is to approximate a large matrix or its inverse using a low rank matrix. Comprehensive reviews can be found in  \citet[Chapter 8.]{Rasmussen2006}, \citet{Quinonero2007},  and \cite{shi2011} for example. An efficient approximation method by adopting a linear random projection has also been proposed by \citet{Banerjee2012}. In our simulations and real case studies, we employed numerical schemes such as using the technique of generalized inverse and converting a non-positive definite matrix to a positive definite matrix by replacing its non-positive eigenvalues with fairly small positive numbers, that is, the so-called {\em jittering}.  Similar strategies have been adopted in the PACE \citep{yao2005functional} as well. For {problems involving large-scale matrix computations}, more advanced low-rank approximation techniques can be integrated into our MCMC steps to further improve scalability.  In terms of the computational scalability, our current MCMC algorithm is able to handle a moderately large dataset with $O(10^4)$ observational points. Alternatively, instead of representing the functional data on grids, we can also adopt basis representations for functional data with an appropriately selected basis system.


\newpage
\begin{acknowledgement}
The authors would like to thank all colleagues in the PO1 project (supported by NIH grant PO1-CA-082710) for collecting the spectroscopy data and the Children's Nutrition Research Center at the Baylor College of Medicine for providing the metabolic data (funded by National Institute of Diabetes and Digestive and Kidney Diseases Grant DK-74387 and the USDA/ARS under Cooperative Agreement 6250-51000-037).  The authors are grateful for all of the comments from the editor, the associate editor, and two referees. Last but not least, the authors greatly appreciate Kirsten Herold of the Writing Lab (School of Public Health, University of Michigan) and Xiaowei Wu (Department of Statistics, Virginia Tech) for their help with editing this article.
 

\end{acknowledgement}

\bibliographystyle{ba}
\bibliography{BAsmooth}

}

\end{document}